\documentstyle[aps,graphicx]{revtex}
 
\def\<{\langle}
\def\>{\rangle}

\begin{document}
\title{Fetching marked items from an unsorted database in NMR ensemble computing}
\author{L. Xiao$^{1,2}$ and G. L. Long$^{1,2,3,4}$}
\address{
$^1$ Department of Physics, Tsinghua University, Beijing 100084,
P R China\\
$^2$ Key Laboratory For Quantum Information and Measurements,
Beijing 100084, P R China\\
$^3$ Institute of Theoretical Physics, Chinese Academy of
Sciences, Beijing 100080, P R China\\
$^4$ Center for Atomic and  Molecular NanoSciences, Tsinghua
University, Beijing 100084, P R China}
\maketitle
\begin{abstract}
Searching a marked item or several marked items from an unsorted database is a very 
difficult mathematical problem.  Using classical computer, it requires
$O(N=2^n)$ steps to find the target. Using a quantum computer, Grover's
algorithm uses $O(\sqrt{N=2^n})$ steps. In NMR ensemble computing, Br\"{u}shweiler's 
algorithm uses $\log N$ steps. In this Letter, we propose an algorithm that fetches marked 
items in an unsorted database directly. It requires only a single query. It can find a 
single marked item or multiple number of items.
\end{abstract}

Unsorted  database search is a very important mathematical problem and it has many 
applications. The problem can be expressed as follows:
Suppose we have a query function  \textbf{f}. \textbf{f} is known to be zero for all 
inputs \textbf{x}, $f(\textbf{x})=0$, except for one or some of the items in a database 
containing $N=2^n$ items, say, $x=z$ and $f(\textbf{z})=1$. The
problem is to find \textbf{z}. On a classical computer one has to
evaluate the function \textbf{f} about $O(2^n)$ times. Using
Grover's algorithm\cite{r1}, we can find \textbf{z} in
$O(2^{n/2})$ steps which is optimal for pure or effective-pure
states\cite{r2}. In contrast, Br\"{u}schweiler's
algorithm finds \textbf{z} in $O(n)$ steps\cite{r3} in an NMR ensemble computer by using 
truly mixed spin states. The speedup is achieved by the  massive
parallelism of representing input states using molecular
sub-ensembles with different spin states. Different spin states perform different 
computations simultaneously in molecules. This parallelism is classical in 
nature\cite{schack}.   In this Letter, we present an algorithm that uses the topological 
nature of the NMR spectrum to actually fetch the marked items from an unsorted database. 
Like the Br\"{u}schweiler algorithm, our algorithm is also for ensemble computers. The 
algorithm is a hybrid algorithm that combines both DNA computing and  quantum computing.

{\bf\it Spin density operator is used to describe NMR ensemble
system}\cite{r4,r4p}.
The usage of polarization operators permits a
concise and suggestive description of spin ensemble quantum
computation. Spin systems containing $n$ spin $I=1/2$ nuclear spins are
chosen to encode the database.  The eigen-state of the nuclear spin systems $|\phi_m\>$
of a weakly-coupled system can be expressed as a direct
product of the single-spin eigen-functions $|\alpha_i\>$,
$|\beta_i\>$ of $I_{iz}$, for instance,
\begin{equation}
|\phi_m\>=|\alpha_1\>|\alpha_2\>|\beta_3\>...|\alpha_{n-1}\>|\beta_n\>
=|\alpha \alpha \beta...\alpha \beta\>=|001...01\>,
\;\;\;\;(m=1,...,2^n),
\end{equation}
where $|\alpha\>=|0\>$ and $|\beta\>=|1\>$. In Liouville space,
the $2^n$ pure states can be expressed by direct products of the
polarization operators $I^{\alpha}$ and $I^{\beta}$,
\begin{equation}
I^{\alpha}=|\alpha\>\<\alpha|=\frac{1}{2}
(\textbf{1}+2I_z)=\left(\begin{array}{cc}
                            1 & 0  \\
                            0 & 0
                    \end{array}\right),
\end{equation}
\begin{equation}
I^{\beta}=|\beta\>\<\beta|=\frac{1}{2}
(\textbf{1}-2I_z)=\left(\begin{array}{cc}
                            0 & 0  \\
                            0 & 1
                    \end{array}\right).
\end{equation}
The corresponding relationship between a density operator $\rho$
and the pure state $|\phi\>$ is
\begin{equation}
|\phi_m\>=|\alpha \alpha \beta ...\alpha \beta\>
\Longleftrightarrow \rho_m=I_1^{\alpha} I_2^{\alpha} I_3^{\beta}
... I_{n-1}^{\alpha} I_n^{\beta}
\end{equation}

Before presenting our algorithm, let's first briefly review the {\bf\it general 
description of a Fourier spectroscopy}\cite{r4}.
Suppose before the acquisition, the density matrix of a spin system is $\rho (0_-)$. In 
order to acquire the wanted information,  a proper radio frequency
(r.f.) pulse such as non-selective r. f. pulse is operated on
$\rho (0_-)$
\begin{equation}
\rho(0_+)=R_y (\beta)\rho(0_-)R_y (\beta)
\end{equation}
where $R_y$ is the rotation operator along $y$ axis.
\begin{equation}
R_y(\beta)=exp\{-i\beta F_y\}
\end{equation}
where the $F_y=\sum\limits I_{ky}$. The subsequent free evolution
is
\begin{equation}
\frac{d\rho(t)}{dt}=-i[H, \rho]-\hat{\hat{\Gamma}}
\{\rho(t)-\rho_0\}
\end{equation}
where $H$ is Hamiltonian and $\hat{\hat{\Gamma}}$ is relax
super-operator. With the Liouville super-operator
$\hat{\hat{L}}=-iH-\hat{\hat{\Gamma}}$ and the equilibrium
density operator $\rho_0$ commuting with the unperturbed
Hamiltonian $H$, The above eq. (7) can be rewritten as
\begin{equation}
\frac{d\rho(t)}{dt}=\hat{\hat{L}}\{\rho(t)-\rho_0\}
\end{equation}
The solution of this equation is
\begin{equation}
\rho(t)=exp(\hat{\hat{L}}t)\{\rho(0_+)-\rho_0\}+\rho_0
\end{equation}
The magnetization $M^+ (t)$ is proportional to the value of the
operator $F^+=F_x+iF_y=\sum\limits I_{kx}+i \sum\limits I_{ky}$
\begin{equation}
M^+ (t)=N\gamma \hbar \<F^+\>(t)=N\gamma \hbar tr\{F^+ \rho(t)\}
\end{equation}
where $N$ is the number spin systems per unit volume.

In a high magnetic field, the Liouville operator $\hat{\hat{L}}$
is invariant under rotation about the $z$-axis and does not mix
components of $\rho$ belonging to different coherence orders $p$,
then a general representation of the complex free induction decay
is
\begin{equation}
M^+ (t)=N\gamma \hbar tr\{F^+ exp(\hat{\hat{L}}t)\rho(0_+)\}
\end{equation}
In the absence of degenerate transitions, each off-diagonal
matrix element of $\rho(t)$ in the eigenbase of $H$ evolves
independently
\begin{equation}
M^+(t)=N\gamma \hbar\sum\limits_{rs} F^+_{sr}
\rho_{rs}(0_+)exp\{(-i\omega_{rs}-\lambda_{rs})t\}=N\gamma
\hbar\sum\limits_{rs} F^+_{rs}
\rho_{sr}(0_+)exp\{(i\omega_{rs}-\lambda_{rs})t\}
\end{equation}
where the transition frequencies are
\begin{equation}
\omega_{rs}=H_{rr}-H_{ss}=\<r|H|r\>-\<s|H|s\>
\end{equation}
and the relaxation rates are
\begin{equation}
\lambda_{rs}=-R_{rs\;rs}= {1 \over {T_2^{(rs)}}}
\end{equation}

 The complex Fourier transformation of the free induction
signal yields the complex spectrum
\begin{equation}
S(\omega)=N\gamma \hbar
\sum\limits_{rs}F^+_{rs}\rho_{sr}(0_+)\frac{1}{i\Delta\omega_{rs}+\lambda_{rs}}
\end{equation}

{\bf \it Obviously, from this formula of complex spectrum, peaks of the spectra have an 
one to one
correspondence to the transitions between energy levels. The states
of an ensemble computer can be identified by transitions between
energy levels in contrast to the common correspondence between
the ensemble-computer states and the energy levels themselves. As we will show later
that peaks of the spectra can be used to label the states of an ensemble
computer}.  A main theme of our algorithm is the use of this correspondence.

 Firstly, we will elucidate the correspondence. The Hamiltonian in a weakly-coupled system 
can be expressed as
\begin{equation}
H=\sum\limits_{j}(\omega_jI_{jz}+\sum\limits_{k\>j}2\pi
J_{jk}I_{jz}I_{kz})\;\;\;\;\;\; (j=1,2,....n)
\end{equation}
The eigen-state $|r\>$ can be expressed as a direct product of
the single-spin eigen-functions
\begin{equation}
|r\>=|i_1i_2...i_{n}\>\;\;\;\;\;\; i_l=0 \;or\; 1
\end{equation}
The eigen-value is
\begin{equation}
\<r|H|r\>=\<i_1i_2...i_{n}|\sum\limits_{j}(\omega_jI_{jz}+\sum\limits_{k\>j}2\pi
J_{jk}I_{jz}I_{kz})|i_1i_2...i_{n}\>=
{1\over 2}\sum\limits_{j}\left((-1)^{i_j}\omega_j+\sum\limits_{k\>j}\pi 
J_{jk}(-1)^{i_j+i_k}\right)
\end{equation}
we adapt the convention $|i_l\>=|1/2\>=|0\>$ corresponding to
excited state (high frequency) and $|i_l\>=|-1/2\>=|1\>$
corresponding to ground state (low frequency). Generally speaking,
in one dimension spectrum, only single-quantum coherence can be
observed directly. The transition selection rules are $\Delta
M=\mp 1$, where $M$ is magnetic quantum number. In our algorithm, we use an aucilla bit 
which is labled as zeroth qubit: $I_0$. The energies of the system when the aucilla bit is 
at 0 and 1 are,
\begin{equation}
\<0, i_1...i_{n}|H|0,
i_1...i_{n}\>=\frac{1}{2}[\omega_0+\sum\limits_{j=1}(-1)^{i_j}\omega_j+
\sum\limits_{k=1}\pi J_{0k}(-1)^{i_k}+\sum\limits_{j}
\sum\limits_{k\>j=1}\pi J_{jk}(-1)^{i_j+i_k}]
\end{equation}
\begin{equation}
\<1, i_1...i_{n}|H|1,
i_1...i_{n}\>=\frac{1}{2}[-\omega_0+\sum\limits_{j=1}(-1)^{i_j}\omega_j-
\sum\limits_{k=1}\pi J_{0k}(-1)^{i_k}+\sum\limits_{j}
\sum\limits_{k\>j=1}\pi J_{jk}(-1)^{i_j+i_k}],
\end{equation}
respectively. If we observe the spectrum of the aucilla bit $I_0$, the transition 
frequencies will be
\begin{equation}
\omega_0+\sum\limits_{k=1}\pi
J_{0k}(-1)^{i_k},\;\;\;\;  \;(k=1,...,n).
\label{omega}
\end{equation}
In the spectrum there are $2^{n}$ peaks. Each peak corresponds to a transition between 
$|0i_1i_2...i_n\>$ and $|1i_1i_2...i_n\>$ in the aucilla bit's spectrum. The frequency of 
the peak is determined by eq.(\ref{omega}). If the state of the aucilla bit is in 0, then 
the transition is from $|0i_1i_2...i_n\>$ to $|1i_1i_2...i_n\>$, and the peak is upward in 
the spectrum. If the state of aucilla bit at the acquisition is at 1, then the transition 
is from $|1i_1i_2...i_n\>$ to $|0i_1i_2...i_n\>$, and the peak is downward. Thus the $2^n$ 
number of peaks in the aucilla bit spectrum correspond to $2^n$ numbers: 
$|i_1i_2...i_n\>$. The upward or downward nature of the peak indicate the aucilla bit's 
state before the acquisition is 0 or 1. The $|i_1i_2...i_n\>=|00...0\>$ state is the far 
left peak(highest frequency) in the spectrum, and $|i_1i_2...i_n\>=|11...1\>$ is the far 
right peak in the spectrum(lowest frequency). In between them are the other states of the 
system.

With the above knowledge in mind, it is easy to present our algorithm: 1) first prepare 
the $n$ bit with the aucilla bit system in the $I^{\alpha}_0$ state. This state actually 
represents $2^n$ items of the database, since
\begin{eqnarray}
I_0^{\alpha}&=&I_0^{\alpha}I_1I_2...I_n=I_0^\alpha(I_1^\alpha+I_1^\beta)(I_2^\alpha+
I_2^\beta)...
(I_n^\alpha+I_n^\beta)
\nonumber\\
&=&I_0^{\alpha}(I_1^\alpha I_2^\alpha...I_n^\alpha+I_1^\alpha I_2^\alpha...I_n^\beta+...
+I_1^\beta I_2^\beta...I_n^\beta);
\end{eqnarray}
2) apply the query function $f$ to the system and store the results of the query in the 
aucilla bit. The query function corresponds to an unitary transformation
\begin{equation}
f=Tr\{UI_0^\alpha\rho_{in}U^\dag I_{0z}\}.
\end{equation}
For truly mixed state,
\begin{equation}
f=\sum\limits_{j=1}^M F(I_0^\alpha \rho_j)=F(\sum\limits_{j=1}^M
I_0^{\alpha}\rho_j); 
\end{equation}
3) measure the aucilla bit's spectrum and read out the results from the spectrum. Marked 
items will be those states with peaks downwards in the spectrum.

We illustrate our idea using one simple
example with $n=2$. The unsorted database can be expressed as (the
extra bit is included in) $\{000\;(I_0^{\alpha}
I_1^{\alpha}I_2^{\alpha}),\;\;001\;(I_0^{\alpha}
I_1^{\alpha}I_2^{\beta}),\;\;010\;(I_0^{\alpha}
I_1^{\beta}I_2^{\alpha}),\;\;011\;(I_0^{\alpha}
I_1^{\beta}I_2^{\beta})\}$. The energy levels and transitions of
the extra bit are:
$
\<000|H|000\>=\frac{1}{2}(\omega_0+\omega_1+\omega_2+\pi
J_{01}+\pi J_{02}+\pi J_{12})$,

$
\<100|H|100\>=-\frac{1}{2}(\omega_0+\omega_1+\omega_2-\pi
J_{01}-\pi J_{02}+\pi J_{12})$,
$\<001|H|001\>=\frac{1}{2}(\omega_0+\omega_1-\omega_2+\pi
J_{01}-\pi J_{02}-\pi J_{12})$,
$\<101|H|101\>=-\frac{1}{2}(\omega_0+\omega_1-\omega_2-\pi
J_{01}+\pi J_{02}-\pi J_{12})$,
$
\<010|H|010\>=\frac{1}{2}(\omega_0-\omega_1+\omega_2-\pi
J_{01}+\pi J_{02}-\pi J_{12})$,
$\<110|H|110\>=-\frac{1}{2}(\omega_0-\omega_1+\omega_2+\pi
J_{01}-\pi J_{02}-\pi J_{12})$,
$\<011|H|011\>=\frac{1}{2}(\omega_0-\omega_1-\omega_2-\pi
J_{01}-\pi J_{02}+\pi J_{12})$,
$\<111|H|111\>=-\frac{1}{2}(\omega_0-\omega_1-\omega_2+\pi
J_{01}+\pi J_{02}+\pi J_{12})$,
$\omega_{04}=\omega_0+ \pi (J_{01}+ J_{02})$,
$\omega_{15}=\omega_0+ \pi (J_{01}- J_{02})$,
$\omega_{26}=\omega_0+ \pi (-J_{01}+ J_{02})$,
$\omega_{37}=\omega_0+ \pi (-J_{01}- J_{02})$.

There are four transitions which corresponding to
$\{00(I_1^\alpha I_2^\alpha),\;\;01(I_1^\alpha
I_2^\beta),\;\;10(I_1^\beta I_2^\alpha),\;\;11(I_1^\beta
I_2^\beta)\}$.  When we acquire spectrum on the state $I_0^\alpha$,
all the peaks in the spectrum are up. After performing the  oracle query $f$, the peaks 
corresponding to marked states will be downwards.

{\bf\it It will be easy to implement the above example in an NMR system.}
The physical system can be the  $^{13}C$ labeled alanine
$^{13}C^1H_3-^{13}C^0H(NH_2^+)-^{13}C^2OOH$. The solvent is
$D_2O$. The $J$ couplings are: $J_{01}=35.1 Hz$ and $J_{02}=54.2 Hz$. Suppose the marked 
items are \textbf{10} and
\textbf{11}.  Starting from  thermal equilibrium
state:
\begin{equation}
\sigma(0_-)=I_z^0+I_z^1+I_z^2,
\end{equation}
state $I_0^\alpha$ is prepared. It can be achieved by
a sequence of selective pulses, non-selective pulse and J-coupling
evolution:
\begin{equation}
(\frac {\pi}{2}){_y^{1,2}} \rightarrow $Grad$
\end{equation}
In which, subscripts denote the directions of the radio frequency.
Superscripts denote the nuclei which the radio frequency is
operated on. $Grad$ refers to gradient field. Then the query function operation is 
performed. It is
\begin{equation}
U=\left[\begin{array}{cccccccc}
1 & 0 & 0 & 0 & 0 & 0 & 0 & 0\\
0 & 1 & 0 & 0 & 0 & 0 & 0 & 0\\
0 & 0 & 0 & 0 & 0 & 0 & 1 & 0\\
0 & 0 & 0 & 0 & 0 & 0 & 0 & 1\\
0 & 0 & 0 & 0 & 1 & 0 & 0 & 0\\
0 & 0 & 0 & 0 & 0 & 1 & 0 & 0\\
0 & 0 & 1 & 0 & 0 & 0 & 0 & 0\\
0 & 0 & 0 & 1 & 0 & 0 & 0 & 0
\end{array}
\right].
\end{equation}
The corresponding pulse sequence is
\begin{equation}
(\frac{\pi}{2})_{\stackrel{-}{y}}^{0} \rightarrow
(\frac{\pi}{2})_{\stackrel{-}{z}}^{0,1} \rightarrow \tau
\rightarrow (\frac{\pi}{2})_{y}^{0}
\end{equation}
where $\tau=\frac{1}{2 J_{01}}$.

We acquire the spectrum of the aucilla bit after the $U$
transformation is operated on the state $I_0^\alpha$. The peaks
down correspond to the marked states \textbf{10},
\textbf{11} respectively. The left one down is
corresponding to \textbf{10}, and the right one down is \textbf{11}.
The spectrum from computer simulation is shown in Fig.1. 

{In summary,} we have proposed an algorithm that finds marked states in an unsorted 
database with a single query. It requires much less computing time and is thus more robust 
against decoherence. In addition, it has several other advantages. The algorithm does not 
require the prior knowledge of the number of marked states in the database. It uses the 
shape of the spectrum, which is a topological property of the spectrum to read out the 
marked states. This makes it more robust to errors and has much less demand on the 
accuracy. As for practical implementation, it is very important to choose the right 
working media. In particular, the aucilla bit should have $J$ coupling with all the other 
bit. 

This work is supported in part by China National Science Foundation, the Fok Ying Tung 
education foundation, major state basic research development program contract no. 
G200077400, the HangTian Science foundation.

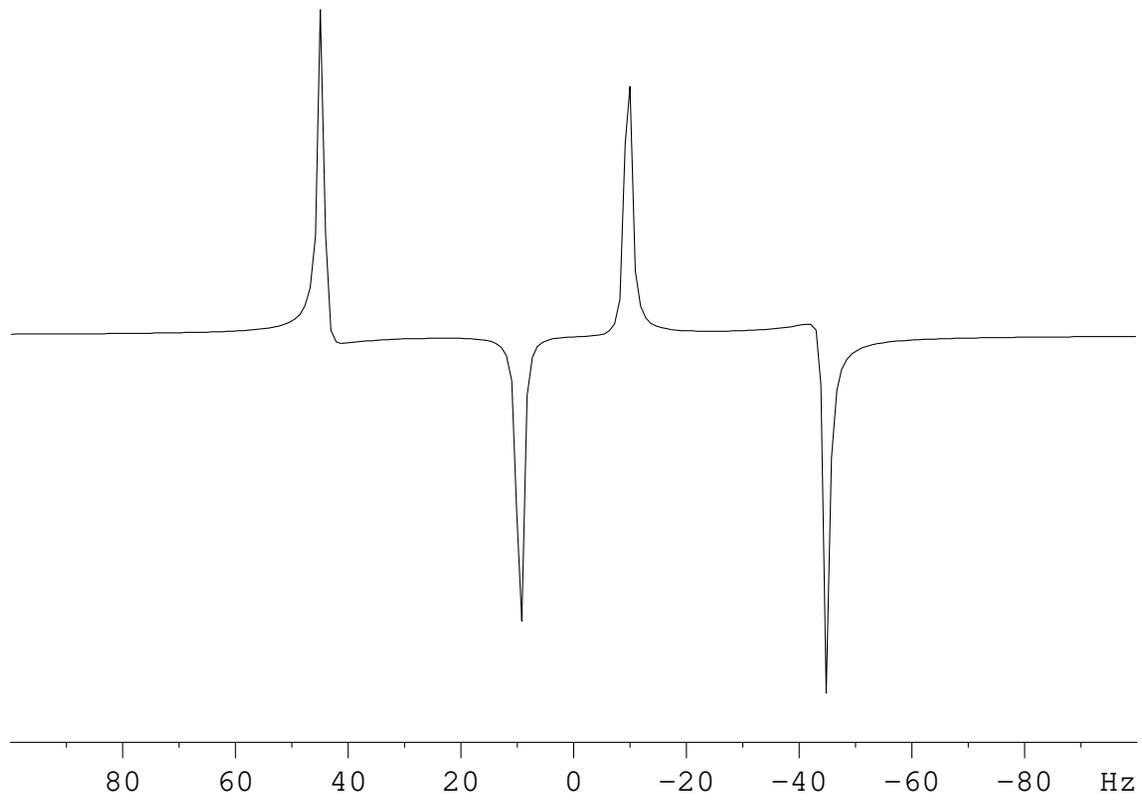
\end{document}